\documentclass[aps, superscriptaddress,reprint,amsmath,amssymb,prl]{revtex4-2}
\usepackage{graphicx}
\usepackage{multirow}
\usepackage{dcolumn}
\usepackage[usenames]{color}
\usepackage{booktabs}
\usepackage[version=4]{mhchem}
\usepackage[utf8]{inputenc}
\usepackage{caption}
\usepackage[colorlinks=true, allcolors=blue]{hyperref}
 
\usepackage{xcolor}
\definecolor{darkgreen}{RGB}{0,150,0}

\usepackage{siunitx}
\DeclareSIUnit\angstrom{\text{Å}}
\DeclareSIUnit{\debye}{D}
\DeclareSIUnit{\calorie}{cal}
\DeclareSIUnit{\kcal}{\kilo\calorie}
\DeclareSIUnit{\step}{step}

\newcommand{\fig}{Fig.~}

\usepackage{booktabs} 

\begin{document}

\title{QCell: Comprehensive Quantum-Mechanical Dataset Spanning \\ Diverse Biomolecular Fragments}

\author{Adil Kabylda$^\parallel$}
\email{adil.kabylda@uni.lu}
\affiliation{Department of Physics and Materials Science, University of Luxembourg, L-1511 Luxembourg City, Luxembourg}

\author{Sergio Suárez-Dou$^\parallel$}
\affiliation{Department of Physics and Materials Science, University of Luxembourg, L-1511 Luxembourg City, Luxembourg}

\author{Nils Davoine}
\affiliation{Department of Physics and Materials Science, University of Luxembourg, L-1511 Luxembourg City, Luxembourg}

\author{Florian N. Br\"unig}
\affiliation{Department of Physics and Materials Science, University of Luxembourg, L-1511 Luxembourg City, Luxembourg}

\author{Alexandre Tkatchenko}
\email{alexandre.tkatchenko@uni.lu}
\affiliation{Department of Physics and Materials Science, University of Luxembourg, L-1511 Luxembourg City, Luxembourg}

\begin{abstract}
\noindent 
Recent advances in machine learning force fields (MLFFs) are revolutionizing molecular simulations by bridging the gap between quantum-mechanical (QM) accuracy and the computational efficiency of mechanistic potentials. However, the development of reliable MLFFs for biomolecular systems remains constrained by the scarcity of high-quality, chemically diverse QM datasets that span all of the major classes of biomolecules expressed in living cells. Crucially, such a comprehensive dataset must be computed using non-empirical or minimally empirical approximations to solving the Schr\"odinger equation. To address these limitations, we introduce the QCell dataset\,--- a curated collection of 525k new QM calculations for biomolecular fragments encompassing carbohydrates, nucleic acids, lipids, dimers, and ion clusters. QCell complements existing datasets, bringing the total number of available data points to 41 million molecular systems, all calculated using hybrid density functional theory with nonlocal many-body dispersion interactions, as captured by the PBE0+MBD(-NL) level of quantum mechanics. The QCell dataset therefore provides a valuable resource for training next-generation MLFFs capable of modeling the intricate interactions that govern biomolecular dynamics beyond small molecules and proteins.
\end{abstract}

\maketitle

\section{Background \& Summary}
The accurate modeling of molecular interactions in (bio)chemical systems has long been a central challenge in computational chemistry and biophysics. Existing methods span a spectrum of approaches that introduce tradeoffs between efficiency and accuracy. At one end are quantum mechanical (QM) methods, ranging from highly accurate techniques such as coupled cluster and quantum Monte Carlo to density functional approximations, which vary from non-empirical to heavily parameterized variants trained on curated datasets. At the other end, empirical atomistic force fields achieve high efficiency through fixed functional forms and parameter sets. These approaches have been invaluable for simulating the structure, dynamics, and function of biomolecules, providing either high accuracy or access to biologically relevant timescales~\cite{hollingsworth2018molecular}. Recently, machine learning force fields (MLFFs) have emerged as a promising alternative, aiming to combine the accuracy of QM methods with the efficiency of classical force fields~\cite{unke2021machine}.

However, successful MLFF applications are critically dependent on the availability of diverse and high-quality QM datasets that faithfully represent the chemical space encountered in (bio)molecular systems~\cite{poltavsky2025crash1, poltavsky2025crash2}. Substantial progress has been made in the development of MLFFs, fueled by datasets such as QM7~\cite{rupp2012fast, ruddigkeit2012enumeration}, QM9~\cite{ramakrishnan2014quantum}, ANI-1~\cite{smith2017ani}, MD17~\cite{chmiela2017machine}, MD22~\cite{chmiela2023accurate}, QM7-X~\cite{hoja2021qm7}, QMugs~\cite{isert2022qmugs}, Splinter~\cite{spronk2023quantum}, SPICE~\cite{eastman2023spice, eastman2024nutmeg}, AQM~\cite{medrano2024dataset}, and AIMNet2~\cite{anstine2024aimnet2}, among many others. These datasets provide extensive coverage for small organic molecules, spanning broad elemental diversity, sizes, conformations, charge and protonation states.

\begin{figure*}
    \centering
    \includegraphics[width=0.87\linewidth]{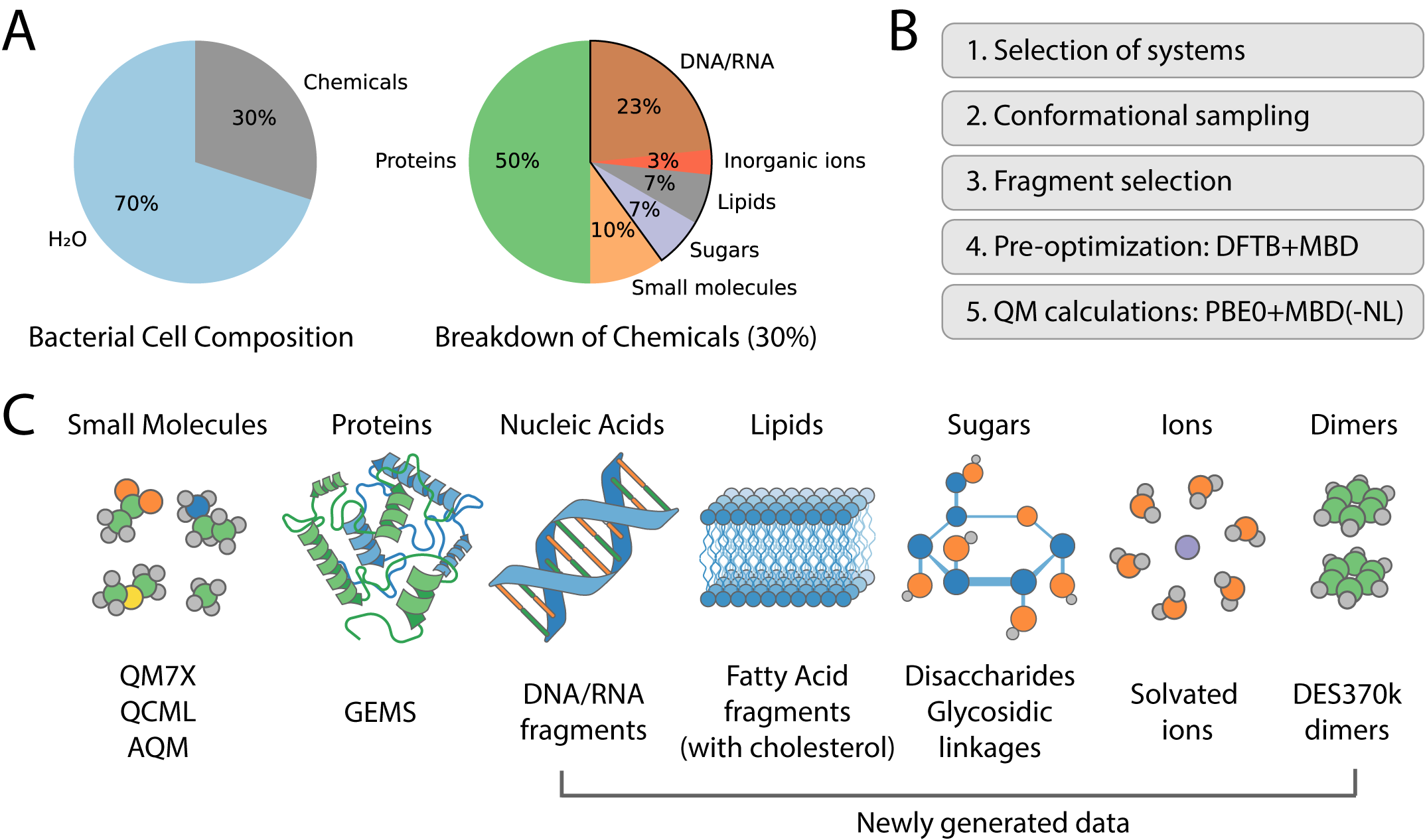}
    \caption{\textbf{Overview.} \textbf{A)} Composition of a bacterial cell by weight, with a breakdown of the chemical constituents~\cite{alberts2022molecular}. About 40\% of these compounds are not properly covered in existing datasets. \textbf{B)} Multi‑step workflow used to construct QCell, beginning with the selection of building blocks, followed by conformational sampling and fragment selection, pre‑optimization with DFTB+MBD, and finally hybrid PBE0+MBD($-\mathrm{NL}$) calculations. \textbf{C)} Coverage of molecular species at the PBE0+MBD(-NL) level of theory, including entries from existing databases and newly generated QCell data for nucleic acid fragments, lipids, sugars, solvated ions, and dimers}
    \label{fig:overview}
\end{figure*}

The GEMS~\cite{unke2024biomolecular}, QCML~\cite{ganscha2025qcml}, and OMol25~\cite{levine2025open} datasets exemplify recent efforts to extend QM coverage across diverse chemical and biomolecular spaces. GEMS employs a hierarchical fragmentation strategy, combining small, transferable protein fragments in gas-phase and aqueous environments with larger, system-specific fragments extending up to 18\,\unit{\angstrom} to capture long-range interactions. QCML systematically maps small-molecule chemical space by enumerating species with up to eight heavy atoms across a wide range of elements and electronic states, providing chemically diverse bonding motifs. OMol25 offers a chemically heterogeneous collection spanning small molecules, biomolecular fragments, metal complexes, and electrolytes; its biomolecular subset includes fragmented protein pockets, gas-phase DNA/RNA fragments, protein–protein and protein–ligand complexes. Despite this progress, existing datasets primarily cover small molecules and protein fragments, leaving significant gaps for three of the four major biomolecular classes, namely nucleic acids, lipids, and carbohydrates, which together constitute roughly 40\% of cellular biomass (Fig.~\ref{fig:overview}A).

Biomolecular chemical space possesses distinct characteristics compared to that of small organic molecules or materials. Instead of vast elemental and topological diversity, biomolecular complexity arises primarily from the conformational space accessible to a relatively limited set of recurring chemical building blocks~\cite{alberts2022molecular}. For instance, proteins are composed primarily of about 20 canonical amino acids, and their intricate functions are dictated by backbone conformations and side-chain rotamer preferences. Similarly, nucleic acids utilize repeating sugar-phosphate backbones and four main nucleobases, with critical conformational variations in sugar pucker and backbone torsions determining their overall structure and interactions. Polysaccharides are formed via various glycosidic linkages between a few monosaccharide types, and their properties depend heavily on the conformations around these linkages. Lipids typically combine a finite set of head groups and fatty acid tails, whose composition and flexibility determine membrane behavior.

In this context, we introduce the QCell dataset, a collection of quantum mechanical data that covers the three major biomolecular classes beyond proteins: lipids, carbohydrates, and nucleic acids, along with relevant ion clusters, water molecules, and non-bonded dimers. The dataset includes 525k newly generated biomolecular fragments, ranging from 2 to 402 atoms, computed at the PBE0+MBD(-NL) level of theory (Fig.~\ref{fig:overview}C). By focusing on fundamental building blocks, the QCell dataset provides an accurate quantum description of the semi-local chemical environments and interaction motifs that recur in larger, more complex biological assemblies.

The chemical element distribution in QCell focuses mainly on biologically relevant elements (H, C, N, O, P, and S) with additional coverage of important biological ions (\ce{Na+}, \ce{K+}, \ce{Cl-}, \ce{Mg^{2+}}, \ce{Ca^{2+}}). This composition provides deeper conformational sampling of the specific chemical environments most relevant to biomolecular systems and allows the QCell dataset to serve as a specialized complement to existing datasets like QCML~\cite{ganscha2025qcml}, QM7-X~\cite{hoja2021qm7}, AQM~\cite{medrano2024dataset}, and GEMS~\cite{unke2024biomolecular}. When combined, they provide extended coverage of chemical space relevant to (bio)molecular simulation, comprising over 41 million data points and spanning 82 chemical elements. The consistent use of the PBE0+MBD(-NL) level of theory across these datasets facilitates their integration into unified training sets for MLFF development. By expanding the coverage to core biomolecular components, QCell enables the development of more comprehensive and transferable MLFFs for applications such as membrane simulations, nucleic acid dynamics, and glycan recognition that were previously limited by the absence of high-level QM data.

\section{Methods}

The QCell dataset was generated using a multi-step workflow (Fig.~\ref{fig:overview}B): (1) curating a library of biomolecular building blocks and generating initial 3D structures; (2) performing extensive conformational sampling using molecular dynamics or dedicated conformer-generation tools; (3) selecting representative fragments from the resulting ensembles; (4) briefly optimizing the selected fragments with the semi-empirical DFTB+MBD method; and (5) running high-quality quantum-mechanical PBE0+MBD(-NL) calculations. Detailed protocols for each molecular class are provided in the subsections below, with scripts available in the accompanying GitHub repository.

\begin{table*}
	\centering
	\caption{\textbf{QCell dataset composition.} Shown alongside existing datasets computed at a consistent PBE0+MBD(-NL) level of theory, enabling direct combination into unified MLFF training sets (QCML~\cite{ganscha2025qcml}, QM7-X~\cite{hoja2021qm7}, AQM~\cite{medrano2024dataset}, GEMS~\cite{unke2024biomolecular}, SPICE~\cite{eastman2023spice, kabylda2025molecular}). Abbreviations: solv. (solvated), bp (base pair), chol. (cholesterol)}
	\label{tab:qcell_table}
	\begin{tabular}{lllllll}
		\toprule
		& \textbf{Type} & \textbf{Size} & \textbf{Atoms} & \textbf{Elements} & \textbf{Theory level} & \textbf{Basis set} \\
        \midrule\midrule
		\multicolumn{6}{l}{\textbf{Small molecules}} \\
		QCML & small molecules & 33,496,171 & 2-36 & 79 elements & PBE0+MBD-NL & tight \\
		QM7-X & small organic molecules & 4,195,237 & 6–23 & H, C, N, O, S, Cl & PBE0+MBD & tight \\
		AQM & drug-like molecules & 59,786 & 2–92 & H, C, N, O, F, P, S, Cl & PBE0+MBD & tight \\
        \midrule
		\multicolumn{6}{l}{\textbf{Proteins}} \\
		GEMS & bottom-up fragments (solv.) & 2,713,986 & 2–120 & H, C, N, O, S & PBE0+MBD & def2-TZVPP \\
		SPICE & dipeptides & 33,849 & 26–60 & H, C, N, O, S & PBE0+MBD & tight \\
		GEMS & top-down fragments (solv.) & 11,819 & 162–321 & H, C, N, O, S & PBE0+MBD & def2-TZVPP \\
        \midrule\midrule
        \multicolumn{6}{l}{\textbf{Nucleic acids}} \\
		QCell & DNA duplex (2 bp, solv.) & 5,333 & 186–246 & H, C, N, O, Na, P & PBE0+MBD-NL & intermediate \\
		QCell & DNA duplex (3 bp, solv.) & 9,534 & 297–382 & H, C, N, O, Na, P & PBE0+MBD-NL & intermediate \\
		QCell & RNA fragments (gas-phase) & 19,971 & 14–282 & H, C, N, O, Na, Mg, S, P & PBE0+MBD-NL & intermediate \\
        \midrule
		\multicolumn{6}{l}{\textbf{Lipids}} \\
	    QCell & fatty acid clusters (1-3-mers) & 12,000 & 125–402 & H, C, N, O, P & PBE0+MBD & intermediate \\
	    QCell & fatty acid clusters (1-2) + chol. & 4,000 & 148–342 & H, C, N, O, P & PBE0+MBD & intermediate \\
        \midrule
		\multicolumn{6}{l}{\textbf{Carbohydrates}} \\
	    QCell & disaccharides & 59,156 & 35–75 & H, C, N, O & PBE0+MBD & tight \\
	    QCell & glycosidic linkages & 14,931 & 38–52 & H, C, N, O & PBE0+MBD & tight \\
        \midrule
        \multicolumn{6}{l}{\textbf{Ions/Water}} \\
		QCell & solvated ions & 25,000 & 4–301 & H, O, Na, Cl, K, Mg, Ca & PBE0+MBD-NL & tight \\
		QCell & water clusters & 5,000 & 6–303 & H, O & PBE0+MBD-NL & tight \\
        \midrule
        \multicolumn{6}{l}{\textbf{Non-covalent dimers}} \\
		QCell & DES370K molecular dimers & 370,956 & 2–34 & 20 elements & PBE0+MBD-NL & tight \\
    	\midrule\midrule
        \multicolumn{2}{l}{\textbf{QCell New}} & 525,881 & 2–402 & 20 elements & PBE0+MBD(-NL) & -- \\
        \multicolumn{2}{l}{\textbf{Total}} & 41,036,729 & 2–402 & 82 elements & PBE0+MBD(-NL) & -- \\
		\bottomrule
	\end{tabular}
\end{table*}

\subsection{Generation of representative fragments}

The current subsection describes steps 1–4 for each molecular class, detailing the specific methods used for initial structure generation, conformational sampling, fragment selection, and pre-optimization.

\textbf{Nucleic Acids.} Solvated double-helical DNA heptamers in canonical A-, B-, and Z-DNA forms~\cite{Neidle1999oxford, Masuda2024} with \ce{Na+} counterions were built using Nucleic Acid Builder~\cite{macke1998modeling} and simulated with the OL21 force field~\cite{zgarbová2021zdna}. The central base-pair triplets covered all base-pair combinations. Each system was equilibrated for 1~ns in NVT, with the temperature ramped from 100~K to 300~K in 10~ps steps, then run for 10~ns in NPT at 300~K.

From the heptamer trajectories, snapshots saved every 100~ps were used to extract central double-stranded trimer fragments. These trimers were then simulated for 10\,ps in NPT at 300\,K with strong positional restraints on nucleotide atoms to relax the surroundings. In addition to trimers, solvated DNA base pair dimers were taken from Ref.~\citenum{berryman2022quantum}, and a subset of smaller gas-phase RNA fragments was taken from OMol25/rna which were processed from BioLiP2~\cite{levine2025open, zhang2024biolip2}.

\textbf{Lipids.} Initial structures of lipid membranes composed of POPC, POPE, POPG, and POPS phospholipids were generated with the CHARMM-GUI Membrane Builder~\cite{wu2014charmm}. These lipids provide a representative set of phospholipid head groups with a palmitoyl–oleoyl–glycerol fatty-acid backbone. To probe sterol–lipid interactions that significantly influence membrane packing and dynamics, we also prepared mixed membranes containing cholesterol: POPC and POPS bilayers were generated at a 3:1 phospholipid-to-cholesterol ratio.

The selected membranes were simulated with the Lipid21 force field~\cite{dickson2022lipid21}. Equilibration involved 20\,ps of NVT at 100\,K followed by 100\,ps of NPT at 300\,K using an anisotropic XY–Z barostat. During equilibration, heavy atoms were restrained with a harmonic potential of 5\,kcal/mol/\unit{\angstrom}$^{2}$. A 500\,ns production simulation was then performed.

The resulting trajectories were sampled randomly over 25000 frames, from which fatty acid monomers, dimers and trimers were selected for subsequent steps (1000 of each n-mer per phospholipid type). Multimers were identified based on geometric proximity: molecules with geometric centers within 5\,\unit{\angstrom} were considered dimers, and those within 6\,\unit{\angstrom} of a dimer were classified as part of a trimer. For cholesterol-containing fragments, only those clusters including at least one cholesterol molecule were retained.

\textbf{Carbohydrates.} A library of 52 common monosaccharides, including both pentose and hexose structures in $\alpha$ and $\beta$ anomeric configurations, was used to construct disaccharides in PyMOL~\cite{PyMOL}. Additionally, we sampled saccharide–peptide linkages, including N-glycosylation involving arginine residues and O-glycosylation involving threonine and serine. These glycosylated residues were capped with ACE and NME groups to mimic peptide termini. In total, 2959 disaccharide structures were generated, representing one unique combination of pentose/hexose and $\alpha$/$\beta$ configuration for each glycosidic linkage and 150 saccharide–peptide molecules.

Conformers were generated with the CREST~\cite{grimme2019exploration, pracht2020automated, pracht2024crest} program, employing a 12\,kcal/mol maximum energy threshold. The resulting ensembles were clustered by linkage dihedral angles, and cluster representatives were selected to ensure broad conformational coverage, retaining at most 100 conformers per amino acid linkage and 20 per disaccharide.

\textbf{Ions and Water.} Solvated ion systems were prepared by placing a single ion at the center of a water box. Bulk water and monovalent ions (\ce{Na+}, \ce{Cl-}, \ce{K+}) in water were simulated in LAMMPS~\cite{thompson2022lammps} under NPT using the MBpol force field implemented in MBX~\cite{zhu2023mbpol, riera2023mbx}. Nosé–Hoover thermostat and barostat were used to maintain the temperature at 298~K and the pressure at 1\,bar, with a time step of 0.5\,fs. Divalent ions (\ce{Ca^{2+}} and \ce{Mg^{2+}}) were simulated for 50\,ns under NPT using the AMBER force field with the TIP3P water model and a time step of 2\,fs~\cite{li2014taking}.

To capture solvation effects across different hydration levels, bulk water and water-ion clusters were cut to contain 1–100 water molecules. Trajectories were sampled every 5\,ps for monovalent ions and bulk water, and every 10\,ps for divalent ions.

\textbf{General MD settings.} All molecular mechanics simulations were carried out using OpenMM~\cite{eastman2024openmm} under NPT conditions at 300\,K, with a 2\,fs time step. A Langevin thermostat was applied with a friction coefficient of 1\,ps$^{-1}$, and pressure was maintained at 1\,atm using a Monte Carlo barostat. For solvated biomolecules, the TIP3P water model~\cite{Sengupta2021para} was used, with \ce{Na+} ions serving as the counterions.

\textbf{Summary and pre-optimization}
Overall, fragments ranged in size from 2 to 402 atoms, with larger fragments chosen to represent important biological motifs such as DNA base-pair stacking and lipid packing interactions. The selected fragments were pre-optimized using the DFTB+MBD method to avoid high-energy clashes~\cite{hourahine2020dftb}. In addition to that, motivated by the importance of dimers in the early stages of developing the general-purpose machine-learned force field SO3LR~\cite{kabylda2025molecular}, we also sourced DES370K dimers for further calculations~\cite{donchev2021quantum}.

\subsection{Quantum Mechanical Calculations}

Within the landscape of electronic structure methods, density functional theory (DFT) offers one of the best tradeoffs between efficiency and accuracy and is widely used for generating large QM datasets. DFT provides a hierarchy of approaches that vary in accuracy and theoretical sophistication, as described by the “Jacob’s ladder”~\cite{perdew2001jacob}. Each rung of Jacob’s ladder represents a higher level of refinement: local density approximation (LDA), generalized gradient approximation (GGA), meta-GGA, hybrid functionals, and advanced formulations such as the random phase approximation or double-hybrid functionals. Ascending the ladder generally improves accuracy but also increases computational cost.

Separate from this hierarchy, functionals also differ in their degree of empiricism. Minimally empirical functionals (such as PBE~\cite{perdew1996generalized} or SCAN~\cite{sun2015strongly}) are constructed from physical constraints with little or no fitting to reference data, while highly empirical functionals are trained on large datasets of experimental or high-level quantum chemical results. Highly empirical methods can achieve impressive accuracy for systems similar to their training data; however, their transferability to new or unseen systems raises concerns. In particular, the reliability of highly empirical functionals in molecular dynamics simulations remains unclear.

Moreover, advanced empirical functionals are typically optimized against coupled cluster data, which is considered a gold standard for small- to medium-sized molecules. However, there is an ongoing debate on how well this method performs for larger systems, where electronic complexity increases. Recently, it has been shown that the most accurate quantum-mechanical methods, CCSD(T) and quantum Monte Carlo, agree for ligand–pocket motifs within 0.5 kcal/mol~\cite{puleva2025extending}, but can struggle to provide consistent reference data for larger molecules or supramolecular complexes with extended $\pi$–$\pi$ interactions. Specifically, the disagreement in the binding energy of a 132-atom buckyball ring complex between the two methods can be up to 12\,kcal/mol~\cite{al2021interactions}. 

\begin{table*}
    \centering
    \caption{\textbf{List of properties stored in the QCell dataset.} The number of Kohn--Sham eigenvalues varies for each molecule. $h_i$ ratios are present only in MBD data, whereas $C_{6}$ and $a_{0}$ ratios appear only in MBD-NL data. File-level information is contained in metadata (\texttt{metadata/free\_atom\_energy} and \texttt{metadata/fhi\_aims\_settings})}

    \label{tab:qcell_properties}
    \begin{tabular}{clllll}
        \toprule
        \textbf{\#} & \textbf{Symbol} & \textbf{Property} & \textbf{HDF5 key} & \textbf{Unit} & \textbf{Shape} \\
        \midrule\midrule

        \multicolumn{6}{l}{\textbf{Structure}} \\
        1  & $Z$ & Atomic numbers   & atomic\_numbers & ---              & (N) \\
        2  & $R$ & Atomic positions & positions       & \unit{\angstrom} & (N, 3) \\
        \midrule

        \multicolumn{6}{l}{\textbf{Energies}} \\
        3  & $E_{\text{tot}}$      & Total energy                         & total\_energy              & \unit{\electronvolt} & () \\
        4  & $E_{\text{form}}$     & Formation energy                     & formation\_energy          & \unit{\electronvolt} & () \\
        \multicolumn{6}{l}{\textit{Total energy components:}} \\
        5  & $\sum_i \varepsilon_i$ & Sum of KS eigenvalues               & sum\_of\_eigenvalues       & \unit{\electronvolt} & () \\
        6  & $\Delta E_{\text{XC}}$ & XC energy correction                & xc\_energy\_correction     & \unit{\electronvolt} & () \\
        7  & $\Delta V_{\text{XC}}$ & XC potential correction             & xc\_potential\_correction  & \unit{\electronvolt} & () \\
        8  & $E_{\text{FA}}$        & Free-atom electrostatic energy      & free\_atoms\_elec          & \unit{\electronvolt} & () \\
        9  & $\Delta E_{\text{H}}$  & Hartree energy correction           & hartree\_correction        & \unit{\electronvolt} & () \\
        10 & $E_{\text{vdW}}$       & van der Waals dispersion energy     & vdw\_energy                & \unit{\electronvolt} & () \\

        \multicolumn{6}{l}{\textit{Derived energy quantities:}} \\
        11 & $E_{\text{kin}}$       & Kinetic energy                      & kinetic\_energy            & \unit{\electronvolt} & () \\
        12 & $E_{\text{elst}}$      & Electrostatic energy                & electrostatic\_energy      & \unit{\electronvolt} & () \\

        \multicolumn{6}{l}{\textit{Decomposition of the XC energy:}} \\
        13 & $E_{\text{HF}}$        & Hartree--Fock energy                & hf\_energy                 & \unit{\electronvolt} & () \\
        14 & $E_{x}$                & Exchange energy                     & x\_energy                  & \unit{\electronvolt} & () \\
        15 & $E_{c}$                & Correlation energy                  & c\_energy                  & \unit{\electronvolt} & () \\
        16 & $E_{\text{XC}}$        & Total XC energy                     & total\_xc\_energy          & \unit{\electronvolt} & () \\
        \midrule

        \multicolumn{6}{l}{\textbf{Forces}} \\
        17 & $F_{\text{tot}}$   & Total forces             & total\_forces            & \unit{\electronvolt\per\angstrom} & (N, 3) \\
        \multicolumn{6}{l}{\textit{Total forces components:}} \\
        18 & $F_{\text{HF}}$    & Hellmann--Feynman forces & hellmann\_feynman\_forces & \unit{\electronvolt\per\angstrom} & (N, 3) \\
        19 & $F_{\text{ion}}$   & Ionic forces             & ionic\_forces            & \unit{\electronvolt\per\angstrom} & (N, 3) \\
        20 & $F_{\text{mult}}$  & Multipole forces         & multipole\_forces        & \unit{\electronvolt\per\angstrom} & (N, 3) \\
        21 & $F_{\text{HFx}}$   & HF exchange forces       & hf\_exchange\_forces     & \unit{\electronvolt\per\angstrom} & (N, 3) \\
        22 & $F_{\text{Pulay}}$ & Pulay+GGA forces         & pulay\_gga\_forces       & \unit{\electronvolt\per\angstrom} & (N, 3) \\
        23 & $F_{\text{vdW}}$   & van der Waals forces     & vdw\_forces              & \unit{\electronvolt\per\angstrom} & (N, 3) \\
        \midrule

        \multicolumn{6}{l}{\textbf{Dipoles and Multipoles}} \\
        24 & $\mu$     & Dipole vector              & dipole                 & \unit{e\times\angstrom}         & (3) \\
        25 & $Q_{\text{tot}}$       & Total quadrupole moment    & quadrupole             & \unit{e\times\angstrom^{2}}     & (3) \\
        26 & $Q_{\text{el}}$        & Electronic quadrupole moment      & electronic\_quadrupole & \unit{e\times\angstrom^{2}}     & (3) \\
        27 & $Q_{\text{ion}}$       & Ionic quadrupole moment         & ionic\_quadrupole      & \unit{e\times\angstrom^{2}}     & (3) \\
        \midrule

        \multicolumn{6}{l}{\textbf{Electronic structure}} \\
        28 & $E_{\text{HOMO}}$      & HOMO energy                & homo\_energy           & \unit{\electronvolt} & () \\
        29 & $E_{\text{LUMO}}$      & LUMO energy                & lumo\_energy           & \unit{\electronvolt} & () \\
        30 & $E_{\text{gap}}$       & HOMO--LUMO gap             & homo\_lumo\_gap        & \unit{\electronvolt} & () \\
        31 & $\{\varepsilon_i\}$    & Kohn–Sham eigenvalues      & ks\_eigenvalues        & \unit{\electronvolt} & (*) \\
        \midrule

        \multicolumn{6}{l}{\textbf{Other / Atomic properties}} \\
        32 & $Q$                    & Total charge               & charge                 & \unit{e}             & () \\
        33 & $h_i$                  & Hirshfeld ratios            & hirshfeld\_ratios      & ---                  & (N) \\
        34 & $C_{6}$     & Atomic $C_{6}$ ratios       & c6\_ratios             & ---                  & (N) \\
        35 & $a_{0}$     & Atomic polarizability ratios & a0\_ratios           & ---                  & (N) \\
        36 & ---                    & Source tag                 & source                 & ---                  & (text) \\
        \bottomrule
    \end{tabular}
\end{table*}

\begin{figure*}
    \centering
    \includegraphics[width=0.9999\linewidth]{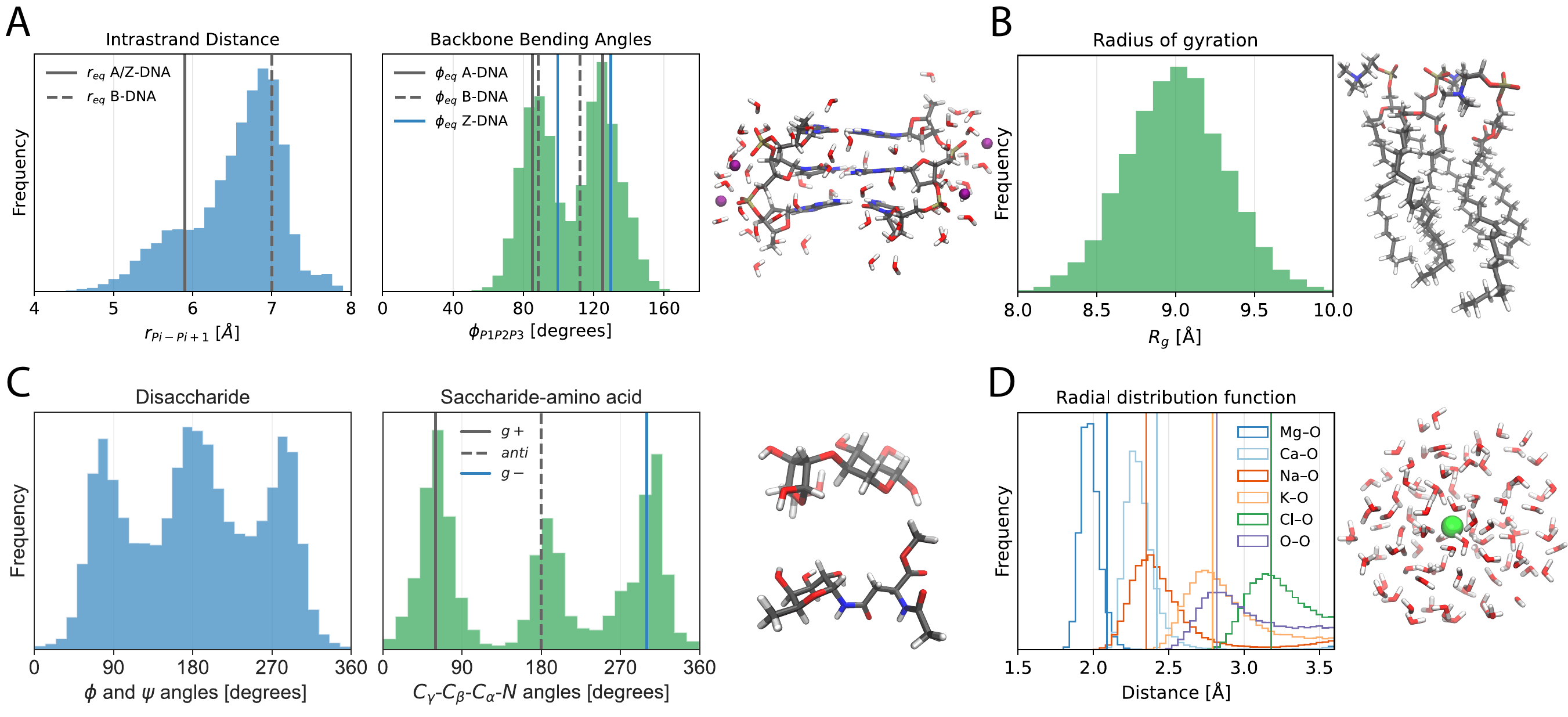}
    \caption{\textbf{Structural distributions across (bio)molecular datasets, with representative structures. A)} Distribution of intra-strand phosphate–phosphate distances (left) and backbone bending angles (right) in DNA trimers. Vertical lines indicate reference values for A-, B-, and Z-DNA~\cite{allemand1998stretched, Mitchell1998}. \textbf{B)} Radius of gyration distribution of fatty acid fragments with more than 300 atoms. \textbf{C)} Distribution of glycosidic linkage dihedrals ($\phi$ and $\psi$) in disaccharides (left) and saccharide–amino acid linkage dihedrals ($C_\gamma$–$C_\beta$–$C_\alpha$–$N$) in glycosylated amino acids (right). Vertical lines indicate gauche$+$, anti, and gauche$-$ conformations. 
    \textbf{D)} Radial distribution functions for ion–oxygen and oxygen–oxygen pairs in ion-water clusters. Vertical lines indicate experimental distances from X-ray and neutron diffraction~\cite{marcus1988ionic, soper2013radial}}
    \label{fig:validation}
\end{figure*}

For this reason, in the current dataset we employed the non-empirical hybrid PBE0 functional with a many-body treatment of dispersion interactions to accurately capture non-covalent interactions~\cite{adamo1999toward, tkatchenko2012accurate, hermann2020density}. Single-point calculations were performed with the FHI-aims code~\cite{blum2009ab, ren2012resolution}. Many-body dispersion was treated using MBD for neutral systems and MBD-NL for ion-containing subsets, as the latter provides improved accuracy for charged species. Scalar-relativistic corrections were included via the atomic ZORA formalism for subsets containing ions. The self-consistent field convergence criteria were set to the following values (or tighter): $10^{-5}$\,eV for the total energy, $10^{-3}$\,eV for the eigenvalue sum, $10^{-5}$ electrons/\unit{\angstrom}$^3$ for the charge density, and $10^{-4}$\,eV/\unit{\angstrom} for the forces. The iteration limit was set to 200, and unconverged calculations were discarded. Calculations used ``tight'' basis sets for small- and medium-sized systems and ``intermediate'' basis sets for fragments with more than 350 atoms, where tight settings were not feasible (see Table~\ref{tab:qcell_table}). Test calculations on 100 lipid fragments (125--268 atoms) comparing tight and intermediate settings yield a force MAE of 0.28~kcal/mol/\AA, well below MLFF training error for this subset (0.6--0.8~kcal/mol/\AA, Fig.~\ref{fig:qcell_mlff}). The consistent use of the PBE0 functional across all datasets ensures that the dominant contribution to energies and forces remains comparable, and the minor variations in basis set and dispersion treatment do not introduce systematic shifts that would compromise MLFF training.

\section{Data Records}

The resulting QCell dataset contains a total of 525,881 QM calculations for biomolecular fragments spanning diverse conformations (Table~\ref{tab:qcell_table}).

The QCell dataset is provided in five HDF5 archive files hosted on a Zenodo data repository and is organized according to the classes listed in Table~\ref{tab:qcell_table} (lipids, carbohydrates, nucleic acids, ions/water, and dimers)~\cite{kabylda2025qcell}. Each molecule in the HDF5 files includes the 34–35 properties listed in Table~\ref{tab:qcell_properties}. A README file is also provided, containing technical usage details and examples illustrating how to access the information stored in the archives (see the \texttt{h5\_to\_extxyz.py} file).

\section{Technical Validation}

To ensure the reliability of the QCell dataset, we validated structural diversity across biomolecular classes using key geometric descriptors (Fig.~\ref{fig:validation}). These analyses serve as diagnostic checks of structural realism and conformational coverage rather than post hoc filters; filtering is enforced earlier in the workflow through the use of well-validated empirical force fields, physically motivated fragment selection, DFTB+MBD pre-optimization, and exclusion of unconverged QM calculations. The goal of validation is therefore to confirm that the final ensembles adequately sample the conformational landscape relevant to each molecular class.

\textbf{Nucleic acids.} For DNA fragments, we analyzed two geometric descriptors that directly reflect the global geometry of double-helical structures, providing a measure of backbone flexibility and helical curvature (\fig~\ref{fig:validation}A). The intra-strand phosphate–phosphate (P--P) distance, $r_{P_i-P_{i+1}}$, is defined as the distance between consecutive phosphate atoms along a single strand. The backbone bending angle, $\phi_{P_1 P_2 P_3}$, is the angle formed by three consecutive phosphate atoms along a strand. Reference values from the literature are $r_{\mathrm{eq}} \approx 5.9$\,\unit{\angstrom} for A/Z-DNA and $r_{\mathrm{eq}} \approx 7.0$\,\unit{\angstrom} for B-DNA for the intra-strand distance, and 80–130$^\circ$ for the backbone bending angle (vertical lines in Fig.~\ref{fig:validation}A)~\cite{allemand1998stretched, Mitchell1998}. The observed distributions confirm that the DNA trimers sample the canonical range of all three helical forms.

\textbf{Lipids.} For lipid fragments, we examined the mass-weighted radius of gyration, $R_g = \sqrt{\sum_i m_i \lVert \mathbf{r}_i - \mathbf{r}_{\mathrm{cm}} \rVert^2 \big/ \sum_i m_i}$, where the sum runs over all atoms $i$ with mass $m_i$ and position $\mathbf{r}_i$, and $\mathbf{r}_{\mathrm{cm}}$ is the center of mass (\fig~\ref{fig:validation}B). $R_g$ provides a measure of overall chain extension and compactness. For fatty-acid fragments with more than 300 atoms, the distribution has a median of 9.0\,\unit{\angstrom} with an interquartile range of 8.8--9.2\,\unit{\angstrom}, reflecting the diversity of packing arrangements.

\textbf{Carbohydrates.} The conformational behavior of carbohydrates is primarily determined by glycosidic torsional angles, which define linkage geometry and flexibility (\fig~\ref{fig:validation}C). Validation confirms that O-glycosidic linkages sample the full torsional range $(0^\circ, 360^\circ]$,  and that N-glycosidic linkages populate all major rotameric basins (gauche$+$, anti, and gauche$-$).

\textbf{Ions/Water.} To assess solvation structure, we computed radial distribution functions (RDFs) between ion--oxygen and oxygen--oxygen pairs, pooled over all cluster sizes (1--100 water molecules) (\fig~\ref{fig:validation}D). For monovalent ions (\ce{Na+}, \ce{K+}, \ce{Cl-}) and water clusters, the first-shell peak positions are in excellent agreement with ion–oxygen distances determined from X-ray and neutron diffraction experiments (2.35\,\unit{\angstrom} for Na--O, 2.79\,\unit{\angstrom} for K--O, 3.18\,\unit{\angstrom} for Cl--O) and the oxygen–oxygen distance in bulk water (2.82\,\unit{\angstrom})~\cite{marcus1988ionic, soper2013radial}. For divalent ions (\ce{Mg^{2+}}, \ce{Ca^{2+}}), the RDF peaks are shifted to shorter distances relative to experiment (2.09\,\unit{\angstrom} for Mg--O, 2.42\,\unit{\angstrom} for Ca--O), likely due to the difficulty of accurately modeling divalent ion solvation with classical force fields~\cite{marcus1988ionic, li2014taking}. Nevertheless, the distributions still capture the correct first- and second-shell structure and remain within physically meaningful ranges.

\begin{figure}
    \centering
    \includegraphics[width=0.9999\linewidth]{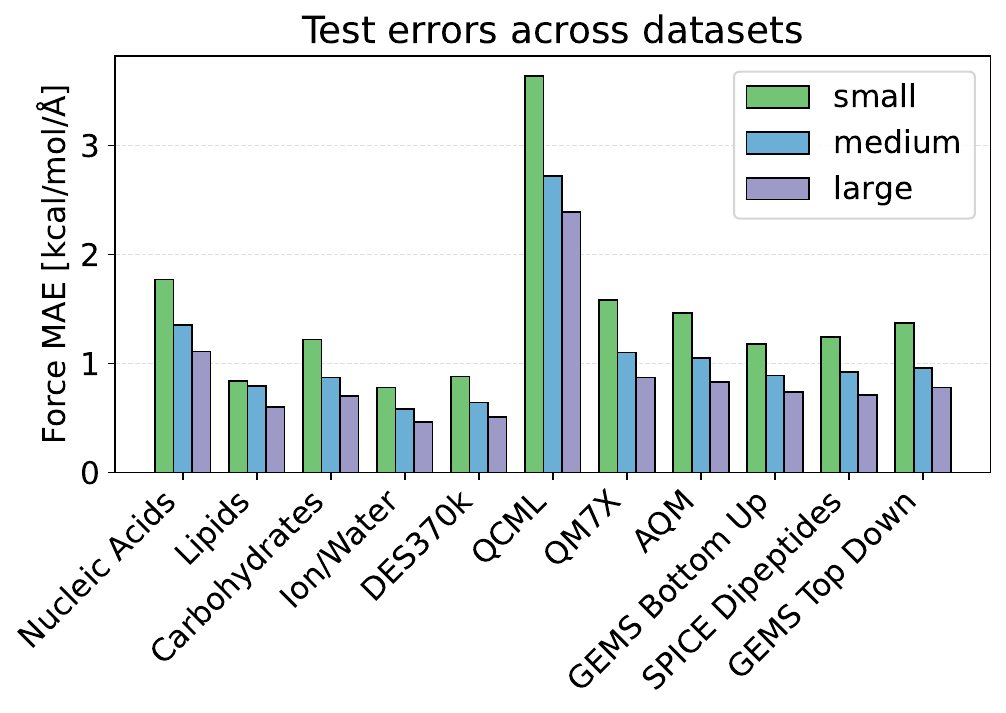}
    \caption{\textbf{Test set errors for machine learning force fields.} Force mean absolute errors [kcal/mol/\unit{\angstrom}] for SO3LR models decrease systematically with model capacity across all subsets, confirming consistent data quality for chemically diverse systems. Small, medium, and large models correspond to feature dimensions $H = 128/256/512$ and message-passing layers $T = 2/3/3$, see Table~\ref{tab:training_params} for more hyperparameters}
    \label{fig:qcell_mlff}
\end{figure}

\textbf{Machine learning models.} To evaluate the dataset in a realistic application, we trained a state-of-the-art machine learning force field on all subsets listed in Table~\ref{tab:qcell_table} and measured its accuracy on held-out test configurations. As a representative model, we employed the SO3LR architecture (based on SO3krates)~\cite{frank2024euclidean, kabylda2025molecular}, which is particularly well suited for biomolecular systems because it explicitly incorporates long-range electrostatic and dispersion interactions, as well as electronic degrees of freedom, and can therefore describe charged and open-shell structures. The model was trained to predict formation energies and atomic forces from atomic numbers, coordinates, and electronic state information (total charge and spin multiplicity). Fig.~\ref{fig:qcell_mlff} summarizes the force mean absolute errors (MAEs) across different molecular classes and model sizes. The errors decrease systematically with increasing model capacity, reaching values below 1\,kcal/mol/\unit{\angstrom} for most subsets. This highlights both the internal consistency of QCell and the ability of modern MLFFs to generalize across chemically diverse systems.

\textbf{MLFF training and parameters.} The SO3LR~\cite{frank2024euclidean, kabylda2025molecular} models were trained using the hyperparameters listed in Table~\ref{tab:training_params}. The small, medium, and large settings form a controlled capacity sweep to quantify how accuracy scales with model size (increasing feature dimension $H$, number of message-passing layers $T$, attention heads $h$, and radial basis functions $k$). Specific values were chosen based on settings validated in prior studies~\cite{kabylda2025molecular, frank2024euclidean}. Training employed a combined loss on forces, dipole moments, Hirshfeld ratios, and energies with weighting factors of 100:10:10:1, using the AdamW optimizer with exponential learning rate decay (factor 0.99 every 1M steps), gradient clipping at global norm 100, and a general robust loss with $\alpha = 1.0$~\cite{barron2019general}. The QCML and QM7-X subsets were sampled 10$\times$ less frequently to ensure balanced training. Long-range interactions used a 10\,\unit{\angstrom} cutoff with electrostatics damping $\sigma = 4$ and dispersion damping $\gamma = 1.2$. All models were trained on A100 GPUs for 180\,GPU-hours.
\pagebreak

\begin{table}[ht]
\centering
\caption{\textbf{Training hyperparameters of the MLFF models}}
\label{tab:training_params}
\begin{tabular}{lccc}
\toprule
\textbf{Parameter} & \textbf{Small} & \textbf{Medium} & \textbf{Large} \\
\midrule\midrule
Cutoff radius (\unit{\angstrom}) & 4.5 & 5.0 & 5.0 \\
Feature dimension ($H$) & 128 & 256 & 512 \\
Message-passing layers ($T$) & 2 & 3 & 3 \\
Number of heads ($h$) & 4 & 8 & 8 \\
Maximal degree ($L$) & 4 & 4 & 4 \\
Radial basis functions ($k$) & 32 & 64 & 128 \\
Batch size ($B$) & 128 & 128 & 64 \\
Learning rate & $5\times10^{-4}$ & $1\times10^{-4}$ & $1\times10^{-4}$ \\
\midrule
\end{tabular}
\end{table}

\section{Conclusion}

We have introduced QCell, a curated collection of 525k quantum-mechanical calculations for biomolecular fragments encompassing nucleic acids, lipids, carbohydrates, solvated ions, and molecular dimers. Existing QM resources have primarily covered small organic molecules and protein fragments, leaving roughly 40\% of cellular biomass underrepresented. QCell begins to address this gap with calculations at the PBE0+MBD(-NL) level of theory, compatible with companion datasets that together yield over 41 million structures spanning 82 chemical elements. Structural validation and benchmark MLFF training confirm broad conformational coverage and consistent data quality across all subsets. By providing high-quality reference data for these underrepresented biomolecular classes, QCell offers a resource for training transferable machine learning force fields capable of modeling biomolecular systems beyond proteins.

\section*{Data Availability}

The QCell dataset has been deposited on Zenodo (\url{https://doi.org/10.5281/zenodo.18385612})~\cite{kabylda2025qcell} and is available as five HDF5 archives (\texttt{qcell\_nucleic\_acids.h5}, \texttt{qcell\_lipids.h5}, \texttt{qcell\_sugars.h5}, \texttt{qcell\_ions\_water.h5}, and \texttt{qcell\_dimers.h5}), a set of preconverted XYZ files with the most relevant properties (\texttt{qcell\_xyz.tar.gz}), conversion script (\texttt{h5\_to\_extxyz.py}), and an example FHI-aims settings file (\texttt{control.in}).

\section*{Code Availability}

Data generation relied on publicly available software: FHI-aims 221103/231212, OpenMM 8.1.1, PyMOL 3.1.0, LAMMPS 06-23-2022, MBX 1.2.0, CREST 3.0.2, DFTB+ 23.1, and CHARMM-GUI (accessed in January 2025). The full data-generation workflow and scripts are available in a public GitHub repository (\url{https://github.com/general-molecular-simulations/qcell}).

\section*{Acknowledgements}

The authors express their gratitude to Raul Ian Sosa and Almaz Khabibrakhmanov for their help with FHI-aims code compilations, and to Joshua T. Berryman, Oliver T. Unke, and Klaus-Robert M\"{u}ller for helpful discussions. The data generation was performed on the Luxembourg national supercomputer MeluXina.

\section*{Funding}

A.K. acknowledges financial support from the Luxembourg National Research Fund (FNR AFR Ph.D. Grant 15720828). S.S.D., N.D., F.B., and A.T. acknowledge the Luxembourg National Research Fund under grant FNR-CORE MBD-in-BMD (18093472) and the European Research Council under ERC-AdG grant FITMOL (101054629). 

\section*{Contributions}

A.K. and S.S.D. contributed equally. A.K. conceived the project. A.K. and S.S.D. selected relevant systems. S.S.D. generated structures for lipids, carbohydrates, and nucleic acids. N.D. and F.B. generated structures for ions and water. A.K. and S.S.D. performed the reference calculations. A.K. trained the ML models. A.K. drafted the manuscript with input from all authors. A.K. and S.S.D. created the figures. All authors discussed the results and contributed to editing the manuscript. A.T. supervised the project. Correspondence to A.K. and A.T.

\twocolumngrid

\bibliography{references}

@article{al2021interactions,
  title={Interactions between large molecules pose a puzzle for reference quantum mechanical methods},
  author={Al-Hamdani, Yasmine S and Nagy, P{\'e}ter R and Zen, Andrea and Barton, Dennis and K{\'a}llay, Mih{\'a}ly and Brandenburg, Jan Gerit and Tkatchenko, Alexandre},
  journal={Nat. Commun.},
  volume={12},
  number={1},
  pages={3927},
  year={2021},
  doi = {10.1038/s41467-021-24119-3}

}

@article{hermann2020density,
  title={Density functional model for van der Waals interactions: Unifying many-body atomic approaches with nonlocal functionals},
  author={Hermann, Jan and Tkatchenko, Alexandre},
  journal={Phys. Rev. Lett.},
  volume={124},
  number={14},
  pages={146401},
  year={2020},
  doi={10.1103/PhysRevLett.124.146401}
}

@article{berryman2022quantum,
    author = {Berryman, Joshua T. and Taghavi, Amirhossein and Mazur, Florian and Tkatchenko, Alexandre},
    title = {Quantum machine learning corrects classical forcefields: Stretching DNA base pairs in explicit solvent},
    journal = {J. Chem. Phys. },
    volume = {157},
    number = {6},
    pages = {064107},
    year = {2022},
    month = {08},
    issn = {0021-9606},
    doi = {10.1063/5.0094727},

}

@article{zhu2023mbpol,
author = {Zhu, Xuanyu and Riera, Marc and Bull-Vulpe, Ethan F. and Paesani, Francesco},
title = {MB-pol(2023): Sub-chemical Accuracy for Water Simulations from the Gas to the Liquid Phase},
journal = {J. Chem. Theory Comput.},
volume = {19},
number = {12},
pages = {3551-3566},
year = {2023},
doi = {10.1021/acs.jctc.3c00326},
}

@article{wu2014charmm,
author = {Wu, Emilia L. and Cheng, Xi and Jo, Sunhwan and Rui, Huan and Song, Kevin C. and Dávila-Contreras, Eder M. and Qi, Yifei and Lee, Jumin and Monje-Galvan, Viviana and Venable, Richard M. and Klauda, Jeffery B. and Im, Wonpil},
title = {CHARMM-GUI Membrane Builder toward realistic biological membrane simulations},
journal = {J. Comput. Chem.},
volume = {35},
number = {27},
pages = {1997-2004},
doi = {https://doi.org/10.1002/jcc.23702},
year = {2014}
}

@book{alberts2022molecular,
  title={Molecular biology of the cell: seventh international student edition with registration card},
  author={Alberts, Bruce and Heald, Rebecca and Johnson, Alexander and Morgan, David and Raff, Martin and Roberts, Keith and Walter, Peter},
  year={2022},
  publisher={WW Norton \& Company}
}

@article{poltavsky2025crash1,
  title={Crash testing machine learning force fields for molecules, materials, and interfaces: model analysis dynamics in the TEA Challenge 2023},
  author={Poltavsky, Igor and Charkin-Gorbulin, Anton and Puleva, Mirela and Fonseca, Gr{\'e}gory and Batatia, Ilyes and Browning, Nicholas J and Chmiela, Stefan and Cui, Mengnan and Frank, J Thorben and Heinen, Stefan and others},
  journal={Chem. Sci.},
  volume={16},
  issue={8},
  pages={3720-3737},
  year={2025},
  publisher={Royal Society of Chemistry},
  doi={10.1039/D4SC06529H}
}

@article{poltavsky2025crash2,
  title={Crash testing machine learning force fields for molecules, materials, and interfaces: molecular dynamics in the TEA Challenge 2023},
  author={Poltavsky, Igor and Charkin-Gorbulin, Anton and Puleva, Mirela and Fonseca, Gr{\'e}gory and Batatia, Ilyes and Browning, Nicholas J and Chmiela, Stefan and Cui, Mengnan and Frank, J Thorben and Heinen, Stefan and others},
  journal={Chem. Sci.},
  volume={16},
  issue={8},
  pages={3738--3754},
  year={2025},
  publisher={Royal Society of Chemistry},
  doi={10.1039/D4SC06530A}
}

@article{levine2025open,
  title={The Open Molecules 2025 (OMol25) Dataset, Evaluations, and Models},
  author={Levine, Daniel S and Shuaibi, Muhammed and Spotte-Smith, Evan Walter Clark and Taylor, Michael G and Hasyim, Muhammad R and Michel, Kyle and Batatia, Ilyes and Cs{\'a}nyi, G{\'a}bor and Dzamba, Misko and Eastman, Peter and others},
  journal={arXiv preprint arXiv:2505.08762},
  year={2025},
  doi={10.48550/arXiv.2505.08762}
}

@article{hollingsworth2018molecular,
  title={Molecular dynamics simulation for all},
  author={Hollingsworth, Scott A and Dror, Ron O},
  journal={Neuron},
  volume={99},
  number={6},
  pages={1129--1143},
  year={2018},
  publisher={Elsevier},
  doi={10.1016/j.neuron.2018.08.011}
}

@article{eastman2024nutmeg,
  title={Nutmeg and SPICE: models and data for biomolecular machine learning},
  author={Eastman, Peter and Pritchard, Benjamin P and Chodera, John D and Markland, Thomas E},
  journal={J. Chem. Theory Comput.},
  volume={20},
  number={19},
  pages={8583--8593},
  year={2024},
  publisher={ACS Publications},
  doi={10.1021/acs.jctc.4c00794}
}

@article{unke2021machine,
	title        = {Machine learning force fields},
	author       = {Unke, Oliver T and Chmiela, Stefan and Sauceda, Huziel E and Gastegger, Michael and Poltavsky, Igor and Schütt, Kristof T and Tkatchenko, Alexandre and Müller, Klaus-Robert},
	year         = 2021,
	journal      = {Chem. Rev.},
	publisher    = {ACS Publications},
	volume       = 121,
	number       = 16,
	pages        = {10142--10186},
        doi={10.1021/acs.chemrev.0c01111}
}

@article{chmiela2017machine,
	title        = {Machine learning of accurate energy-conserving molecular force fields},
	author       = {Chmiela, Stefan and Tkatchenko, Alexandre and Sauceda, Huziel E and Poltavsky, Igor and Sch{\"u}tt, Kristof T and M{\"u}ller, Klaus-Robert},
	year         = 2017,
	journal      = {Sci. Adv.},
	publisher    = {American Association for the Advancement of Science},
	volume       = 3,
	number       = 5,
	pages        = {e1603015},
        doi          = {10.1126/sciadv.1603015}
}

@article{eastman2023spice,
  title={Spice, a dataset of drug-like molecules and peptides for training machine learning potentials},
  author={Eastman, Peter and Behara, Pavan Kumar and Dotson, David L and Galvelis, Raimondas and Herr, John E and Horton, Josh T and Mao, Yuezhi and Chodera, John D and Pritchard, Benjamin P and Wang, Yuanqing and others},
  journal={Sci. Data},
  volume={10},
  number={1},
  pages={11},
  year={2023},
  publisher={Nature Publishing Group UK London},
  doi={10.1038/s41597-022-01882-6}
}

@article{hoja2021qm7,
  title={QM7-X, a comprehensive dataset of quantum-mechanical properties spanning the chemical space of small organic molecules},
  author={Hoja, Johannes and Medrano Sandonas, Leonardo and Ernst, Brian G and Vazquez-Mayagoitia, Alvaro and DiStasio Jr, Robert A and Tkatchenko, Alexandre},
  journal={Sci. Data},
  volume={8},
  number={1},
  pages={43},
  year={2021},
  publisher={Nature Publishing Group UK London},
  doi={10.1038/s41597-021-00812-2}
}

@article{donchev2021quantum,
  title={Quantum chemical benchmark databases of gold-standard dimer interaction energies},
  author={Donchev, Alexander G and Taube, Andrew G and Decolvenaere, Elizabeth and Hargus, Cory and McGibbon, Robert T and Law, Ka-Hei and Gregersen, Brent A and Li, Je-Luen and Palmo, Kim and Siva, Karthik and others},
  journal={Sci. Data},
  volume={8},
  number={1},
  pages={55},
  year={2021},
  publisher={Nature Publishing Group UK London},
  doi={10.1038/s41597-021-00833-x}
}

@article{medrano2024dataset,
  title={Dataset for quantum-mechanical exploration of conformers and solvent effects in large drug-like molecules},
  author={Medrano Sandonas, Leonardo and Van Rompaey, Dries and Fallani, Alessio and Hilfiker, Mathias and Hahn, David and Perez-Benito, Laura and Verhoeven, Jonas and Tresadern, Gary and Kurt Wegner, Joerg and Ceulemans, Hugo and others},
  journal={Sci. Data},
  volume={11},
  number={1},
  pages={742},
  year={2024},
  publisher={Nature Publishing Group UK London},
  doi={10.1038/s41597-024-03521-8}
}

@article{ramakrishnan2014quantum,
  title={Quantum chemistry structures and properties of 134 kilo molecules},
  author={Ramakrishnan, Raghunathan and Dral, Pavlo O and Rupp, Matthias and Von Lilienfeld, O Anatole},
  journal={Sci. data},
  volume={1},
  number={1},
  pages={1--7},
  year={2014},
  publisher={Nature Publishing Group},
  doi={10.1038/sdata.2014.22}
}

@article{isert2022qmugs,
  title={QMugs, quantum mechanical properties of drug-like molecules},
  author={Isert, Clemens and Atz, Kenneth and Jim{\'e}nez-Luna, Jos{\'e} and Schneider, Gisbert},
  journal={Sci. Data},
  volume={9},
  number={1},
  pages={273},
  year={2022},
  publisher={Nature Publishing Group UK London},
  doi={10.1038/s41597-024-04206-y}
}

@article{smith2017ani,
  title={ANI-1, A data set of 20 million calculated off-equilibrium conformations for organic molecules},
  author={Smith, Justin S and Isayev, Olexandr and Roitberg, Adrian E},
  journal={Sci. Data},
  volume={4},
  number={1},
  pages={1--8},
  year={2017},
  publisher={Nature Publishing Group},
  doi={10.1038/sdata.2017.193}
}

@article{frank2024euclidean,
  title={A Euclidean transformer for fast and stable machine learned force fields},
  author={Frank, J Thorben and Unke, Oliver T and M{\"u}ller, Klaus-Robert and Chmiela, Stefan},
  journal={Nat. Commun.},
  volume={15},
  number={1},
  pages={6539},
  year={2024},
  publisher={Nature Publishing Group UK London},
  doi={10.1038/s41467-024-50620-6}
}

@article{dickson2022lipid21,
  title={Lipid21: Complex Lipid Membrane Simulations with AMBER},
  author={Dickson, Callum J and Walker, Ross C and Gould, Ian R},
  journal={J. Chem. Theory Comput.},
  volume={18},
  number={3},
  pages={1726--1736},
  year={2022},
  publisher={ACS Publications},
  doi={10.1021/acs.jctc.1c01217}
}

@article{chmiela2023accurate,
  title={Accurate global machine learning force fields for molecules with hundreds of atoms},
  author={Chmiela, Stefan and Vassilev-Galindo, Valentin and Unke, Oliver T and Kabylda, Adil and Sauceda, Huziel E and Tkatchenko, Alexandre and M{\"u}ller, Klaus-Robert},
  journal={Sci. Adv.},
  volume={9},
  number={2},
  pages={eadf0873},
  year={2023},
  publisher={American Association for the Advancement of Science},
  doi={10.1126/sciadv.adf0873}
}

@article{unke2024biomolecular,
  title={Biomolecular dynamics with machine-learned quantum-mechanical force fields trained on diverse chemical fragments},
  author={Unke, Oliver T and St{\"o}hr, Martin and Ganscha, Stefan and Unterthiner, Thomas and Maennel, Hartmut and Kashubin, Sergii and Ahlin, Daniel and Gastegger, Michael and Medrano Sandonas, Leonardo and Berryman, Joshua T and others},
  journal={Sci. Adv.},
  volume={10},
  number={14},
  pages={eadn4397},
  year={2024},
  publisher={American Association for the Advancement of Science},
  doi={10.1126/sciadv.adn4397}
}

@article{anstine2024aimnet2,
  title={AIMNet2: a neural network potential to meet your neutral, charged, organic, and elemental-organic needs},
  author={Anstine, Dylan M and Zubatyuk, Roman and Isayev, Olexandr},
  journal={Chem. Sci.},
  volume={16},
  number={23},
  pages={10228--10244},
  year={2025},
  publisher={Royal Society of Chemistry},
  doi={10.1039/D4SC08572H}
}

@article{blum2009ab,
  title={Ab initio molecular simulations with numeric atom-centered orbitals},
  author={Blum, Volker and Gehrke, Ralf and Hanke, Felix and Havu, Paula and Havu, Ville and Ren, Xinguo and Reuter, Karsten and Scheffler, Matthias},
  journal={Comput. Phys. Commun.},
  volume={180},
  number={11},
  pages={2175--2196},
  year={2009},
  publisher={Elsevier},
  doi={10.1016/j.cpc.2009.06.022}
}

@article{ren2012resolution,
  title={Resolution-of-identity approach to Hartree--Fock, hybrid density functionals, RPA, MP2 and GW with numeric atom-centered orbital basis functions},
  author={Ren, Xinguo and Rinke, Patrick and Blum, Volker and Wieferink, J{\"u}rgen and Tkatchenko, Alexandre and Sanfilippo, Andrea and Reuter, Karsten and Scheffler, Matthias},
  journal={New J. Phys.},
  volume={14},
  number={5},
  pages={053020},
  year={2012},
  publisher={IOP Publishing},
  doi={10.1088/1367-2630/14/5/053020}
}

@article{kabylda2025molecular,
  title={Molecular Simulations with a Pretrained Neural Network and Universal Pairwise Force Fields},
  author={Kabylda, Adil and Frank, J Thorben and Su{\'a}rez-Dou, Sergio and Khabibrakhmanov, Almaz and Medrano Sandonas, Leonardo and Unke, Oliver T and Chmiela, Stefan and M{\"u}ller, Klaus-Robert and Tkatchenko, Alexandre},
  journal={J. Am. Chem. Soc.},
  volume={147},
  number={37},
  pages={33723},
  year={2025},
  doi={10.1021/jacs.5c09558}
}

@article{soper2013radial,
  title={The radial distribution functions of water as derived from radiation total scattering experiments: is there anything we can say for sure?},
  author={Soper, Alan K},
  journal={Int. Sch. Res. Notices},
  volume={2013},
  number={1},
  pages={279463},
  year={2013},
  publisher={Wiley Online Library},
  doi={10.1155/2013/279463}
}

@article{ganscha2025qcml,
  title={The QCML dataset, Quantum chemistry reference data from 33.5 M DFT and 14.7 B semi-empirical calculations},
  author={Ganscha, Stefan and Unke, Oliver T and Ahlin, Daniel and Maennel, Hartmut and Kashubin, Sergii and M{\"u}ller, Klaus-Robert},
  journal={Sci. Data},
  volume={12},
  number={1},
  pages={406},
  year={2025},
  publisher={Nature Publishing Group UK London},
  doi={10.1038/s41597-025-04720-7}
}

@article{thompson2022lammps,
  title={LAMMPS-a flexible simulation tool for particle-based materials modeling at the atomic, meso, and continuum scales},
  author={Thompson, Aidan P and Aktulga, H Metin and Berger, Richard and Bolintineanu, Dan S and Brown, W Michael and Crozier, Paul S and In't Veld, Pieter J and Kohlmeyer, Axel and Moore, Stan G and Nguyen, Trung Dac and others},
  journal={Comput. Phys. Commun.},
  volume={271},
  pages={108171},
  year={2022},
  publisher={Elsevier},
  doi = {10.1016/j.cpc.2021.108171}, 
}

@unpublished{PyMOL,
	Author = {{Schr\"odinger, LLC}},
	Month = {November},
	Title = {The {PyMOL} Molecular Graphics System, Version~1.8},
	Year = {2015}
}

@inbook{macke1998modeling,
author = {Macke, Thomas J. and Case, David A.},
title = {Modeling Unusual Nucleic Acid Structures},
booktitle = {Molecular Modeling of Nucleic Acids},
chapter = {24},
pages = {379-393},
doi = {10.1021/bk-1998-0682.ch024},
publisher = {American Chemical Society},
year = {1997}
}

@Article{pracht2020automated,
author ="Pracht, Philipp and Bohle, Fabian and Grimme, Stefan",
title  ="Automated exploration of the low-energy chemical space with fast quantum chemical methods",
journal  ="Phys. Chem. Chem. Phys.",
year  ="2020",
volume  ="22",
issue  ="14",
pages  ="7169-7192",
publisher  ="The Royal Society of Chemistry",
doi  ="10.1039/C9CP06869D",
}

@book{Neidle1999oxford,
    author = {Neidle, Stephen},
    title = {Oxford Handbook of Nucleic Acid Structure},
    publisher = {Oxford University Press},
    year = {1999},
    month = {03},
    isbn = {9780198500384},
    doi = {10.1093/oso/9780198500384.001.0001},
}

@article{eastman2024openmm,
  title={OpenMM 8: molecular dynamics simulation with machine learning potentials},
  author={Eastman, Peter and Galvelis, Raimondas and Pel{\'a}ez, Ra{\'u}l P and Abreu, Charlles RA and Farr, Stephen E and Gallicchio, Emilio and Gorenko, Anton and Henry, Michael M and Hu, Frank and Huang, Jing and others},
  journal={J. Phys. Chem. B.},
  volume={128},
  number={1},
  pages={109--116},
  year={2024},
  publisher={ACS Publications},
  doi = {10.1021/acs.jpcb.3c06662},
}

@article{zgarbová2021zdna,
author = {Zgarbová, Marie and Šponer, Jiří and Jurečka, Petr},
title = {Z-DNA as a Touchstone for Additive Empirical Force Fields and a Refinement of the Alpha/Gamma DNA Torsions for AMBER},
journal = {J. Chem. Theory Comput.},
volume = {17},
number = {10},
pages = {6292-6301},
year = {2021},
doi = {10.1021/acs.jctc.1c00697},

}

@article{Sengupta2021para,
author = {Sengupta, Arkajyoti and Li, Zhen and Song, Lin Frank and Li, Pengfei and Merz, Kenneth M. Jr.},
title = {Parameterization of Monovalent Ions for the OPC3, OPC, TIP3P-FB, and TIP4P-FB Water Models},
journal = {J. Chem. Inf. Model.},
volume = {61},
number = {2},
pages = {869-880},
year = {2021},
doi = {10.1021/acs.jcim.0c01390},
}

@Article{Masuda2024,
author={Masuda, Kairi
and Abdullah, Adib A.
and Pflughaupt, Patrick
and Sahakyan, Aleksandr B.},
title={Quantum mechanical electronic and geometric parameters for DNA k-mers as features for machine learning},
journal={Sci. Data},
year={2024},
month={Aug},
day={22},
volume={11},
number={1},
pages={911},
issn={2052-4463},
doi={10.1038/s41597-024-03772-5},
url={https://doi.org/10.1038/s41597-024-03772-5}
}

@article{hourahine2020dftb,
  title={DFTB+, a software package for efficient approximate density functional theory based atomistic simulations},
  author={Hourahine, Ben and Aradi, B{\'a}lint and Blum, Volker and Bonafe, Frank and Buccheri, Alex and Camacho, Cristopher and Cevallos, Caterina and Deshaye, MY and Dumitric{\u{a}}, T and Dominguez, A and others},
  journal={J. Chem. Phys.},
  volume={152},
  number={12},
  pages = {124101},
  year={2020},
  publisher={AIP Publishing},
  doi = {10.1063/1.5143190}
}

@article{pracht2024crest,
  title={CREST—A program for the exploration of low-energy molecular chemical space},
  author={Pracht, Philipp and Grimme, Stefan and Bannwarth, Christoph and Bohle, Fabian and Ehlert, Sebastian and Feldmann, Gereon and Gorges, Johannes and M{\"u}ller, Marcel and Neudecker, Tim and Plett, Christoph and others},
  journal={J. Chem. Phys.},
  volume={160},
  number={11},
  pages = {114110},
  year={2024},
  publisher={AIP Publishing},
  doi = {10.1063/5.0197592}
}

@article{grimme2019exploration,
  title={Exploration of chemical compound, conformer, and reaction space with meta-dynamics simulations based on tight-binding quantum chemical calculations},
  author={Grimme, Stefan},
  journal={J. Chem. Theory Comput.},
  volume={15},
  number={5},
  pages={2847--2862},
  year={2019},
  publisher={ACS Publications},
  doi = {10.1021/acs.jctc.9b00143}
}

@article{allemand1998stretched,
  title={Stretched and overwound DNA forms a Pauling-like structure with exposed bases},
  author={Allemand, Jean F and Bensimon, David and Lavery, Richard and Croquette, Vincent},
  journal={Proc. Natl. Acad. Sci.},
  volume={95},
  number={24},
  pages={14152--14157},
  year={1998},
  publisher={The National Academy of Sciences},
  doi = {10.1073/pnas.95.24.14152},
}

@article{Mitchell1998,
author={Mitchell, Alison},
title={The A to Z of DNA},
journal={Nature},
year={1998},
month={Dec},
day={01},
volume={396},
number={6711},
pages={524-524},
issn={1476-4687},
doi={10.1038/25014},
}

@article{zhang2024biolip2,
  title={BioLiP2: an updated structure database for biologically relevant ligand--protein interactions},
  author={Zhang, Chengxin and Zhang, Xi and Freddolino, Lydia and Zhang, Yang},
  journal={Nucleic Acids Res.},
  volume={52},
  number={D1},
  pages={D404--D412},
  year={2024},
  publisher={Oxford University Press},
  doi={10.1093/nar/gkad630}
}

@article{puleva2025extending,
  title={Extending quantum-mechanical benchmark accuracy to biological ligand-pocket interactions},
  author={Puleva, Mirela and Medrano Sandonas, Leonardo and L{\H{o}}rincz, Bal{\'a}zs D and Charry, Jorge and Rogers, David M and Nagy, P{\'e}ter R and Tkatchenko, Alexandre},
  journal={Nat. Commun.},
  volume={16},
  number={1},
  pages={8583},
  year={2025},
  publisher={Nature Publishing Group UK London},
  doi={10.1038/s41467-025-63587-9}
}

@article{spronk2023quantum,
  title={A quantum chemical interaction energy dataset for accurately modeling protein-ligand interactions},
  author={Spronk, Steven A and Glick, Zachary L and Metcalf, Derek P and Sherrill, C David and Cheney, Daniel L},
  journal={Sci. Data},
  volume={10},
  number={1},
  pages={619},
  year={2023},
  publisher={Nature Publishing Group UK London},
  doi={10.1038/s41597-023-02443-1}
}

@inproceedings{perdew2001jacob,
  title={Jacob’s ladder of density functional approximations for the exchange-correlation energy},
  author={Perdew, John P and Schmidt, Karla},
  booktitle={AIP Conf. Proc.},
  volume={577},
  pages={1--20},
  year={2001},
  organization={American Institute of Physics},
  doi={10.1063/1.1390175}
}

@misc{kabylda2025qcell,
  author       = {Kabylda, Adil and
                  Suárez-Dou, Sergio and
                  Davoine, Nils and
                  Brünig, Florian and
                  Tkatchenko, Alexandre},
  title        = {QCell: Comprehensive Quantum-Mechanical Dataset
                  Spanning Diverse Biomolecular Fragments},
  year         = {2026},
  month        = {jan},
  publisher    = {Zenodo},
  doi          = {10.5281/zenodo.18385612},
  url          = {https://doi.org/10.5281/zenodo.18385612},
  note         = {Dataset on Zenodo}
}

@article{marcus1988ionic,
  title={Ionic radii in aqueous solutions},
  author={Marcus, Yizhak},
  journal={Chem. Rev.},
  volume={88},
  number={8},
  pages={1475},
  year={1988},
  publisher={ACS Publications},
  doi={10.1021/cr00090a003}
}

@inproceedings{barron2019general,
  title={A general and adaptive robust loss function},
  author={Barron, Jonathan T},
  booktitle={Proc. IEEE/CVF Conf. Comput. Vis. Pattern Recognit.},
  pages={4331--4339},
  year={2019},
  url={https://openaccess.thecvf.com/content_CVPR_2019/html/Barron_A_General_and_Adaptive_Robust_Loss_Function_CVPR_2019_paper.html}
}

@article{tkatchenko2012accurate,
  title={Accurate and efficient method for many-body van der Waals interactions},
  author={Tkatchenko, Alexandre and DiStasio Jr, Robert A and Car, Roberto and Scheffler, Matthias},
  journal={Phys. Rev. Lett.},
  volume={108},
  number={23},
  pages={236402},
  year={2012},
  publisher={APS},
  doi={10.1103/PhysRevLett.108.236402}
}

@article{adamo1999toward,
  title={Toward reliable density functional methods without adjustable parameters: The PBE0 model},
  author={Adamo, Carlo and Barone, Vincenzo},
  journal={J. Chem. Phys.},
  volume={110},
  number={13},
  pages={6158--6170},
  year={1999},
  publisher={American Institute of Physics},
  doi={10.1063/1.478522}
}

@article{perdew1996generalized,
  title={Generalized gradient approximation made simple},
  author={Perdew, John P and Burke, Kieron and Ernzerhof, Matthias},
  journal={Phys. Rev. Lett.},
  volume={77},
  number={18},
  pages={3865},
  year={1996},
  publisher={APS},
  doi={10.1103/PhysRevLett.77.3865}
}

@article{sun2015strongly,
  title={Strongly constrained and appropriately normed semilocal density functional},
  author={Sun, Jianwei and Ruzsinszky, Adrienn and Perdew, John P},
  journal={Phys. Rev. Lett.},
  volume={115},
  number={3},
  pages={036402},
  year={2015},
  publisher={APS},
  doi={10.1103/PhysRevLett.115.036402}
}

@article{li2014taking,
  title={Taking into account the ion-induced dipole interaction in the nonbonded model of ions},
  author={Li, Pengfei and Merz Jr, Kenneth M},
  journal={J. Chem. Theory Comput.},
  volume={10},
  number={1},
  pages={289--297},
  year={2014},
  publisher={ACS Publications},
  doi = {10.1021/ct400751u}
}

@article{riera2023mbx,
  title={MBX: A many-body energy and force calculator for data-driven many-body simulations},
  author={Riera, Marc and Knight, Christopher and Bull-Vulpe, Ethan F and Zhu, Xuanyu and Agnew, Henry and Smith, Daniel GA and Simmonett, Andrew C and Paesani, Francesco},
  journal={J. Chem. Phys.},
  volume={159},
  number={5},
  pages={054802},
  year={2023},
  publisher={AIP Publishing},
  doi={10.1063/5.0156036}
}

@article{rupp2012fast,
  title={Fast and accurate modeling of molecular atomization energies with machine learning},
  author={Rupp, Matthias and Tkatchenko, Alexandre and M{\"u}ller, Klaus-Robert and Von Lilienfeld, O Anatole},
  journal={Phys. Rev. Lett.},
  volume={108},
  number={5},
  pages={058301},
  year={2012},
  publisher={APS},
  doi={10.1103/PhysRevLett.108.058301}
}

@article{ruddigkeit2012enumeration,
  title={Enumeration of 166 billion organic small molecules in the chemical universe database GDB-17},
  author={Ruddigkeit, Lars and Van Deursen, Ruud and Blum, Lorenz C and Reymond, Jean-Louis},
  journal={J. Chem. Inf. Model.},
  volume={52},
  number={11},
  pages={2864},
  year={2012},
  publisher={ACS Publications},
  doi={10.1021/ci300415d}
}

\clearpage
\renewcommand{\thesection}{S\arabic{section}}  
\renewcommand{\thetable}{S\arabic{table}}  
\renewcommand{\thefigure}{S\arabic{figure}}
\setcounter{figure}{0}
\setcounter{table}{0}
\onecolumngrid
\end{document}